\documentclass[prl,aps,amsmath,twocolumn,floatfix,showpacs]{revtex4}

\usepackage{epsfig}

\newcommand{\ket}[1]{|#1\rangle}

\newcommand{\mean}[1]{\langle#1\rangle}
\newcommand{\be}{\begin{equation}}
\newcommand{\ee}{\end{equation}}

\begin{document}

\title{Sub-wavelength resolution of optical fields probed by single trapped ions:\\
Interference, phase modulation, and which-way information}

\author{J{\"u}rgen Eschner}

\affiliation{Institut f\"ur Experimentalphysik, Universit\"at Innsbruck,
Technikerstr.~25, 6020 Innsbruck, Austria\\
E-mail: Juergen.Eschner@uibk.ac.at; Homepage:
http://heart-c704.uibk.ac.at/Welcome.html}


\date{\today}

\begin{abstract}

Taking recent experiments as examples, we discuss the conditions for
sub-wavelength probing of optical field structures by single trapped atoms.
We calculate the achievable resolution, highlighting its connection to the
fringe visibility in an interference experiment. We show that seemingly
different physical pictures, such as spatial averaging, phase modulation, and
which-way information, describe the situation equally and lead to identical
results. The connection to Bohr's moving slit experiment is pointed out.

\end{abstract}


\maketitle

\section{Introduction}

In several recent experiments, single trapped ions have been used to map
optical fields, and a resolution considerably below the wavelength has been
reported \cite{Eschner2001,Guthoehrlein2001,Mundt2002}. In the first of these
experiments \cite{Eschner2001}, a part of the resonance fluorescence of a
single Ba$^+$ ion was back-reflected onto the ion with a distant mirror, and
the resulting emission was found to be modulated upon variation of the
ion-mirror distance. In the second experiment \cite{Guthoehrlein2001}, a
single Ca$^+$ ion was placed inside an optical cavity, and light coupled into
the cavity was resonantly scattered. When the ion was shifted along the
cavity axis, periodic variation of the emission was observed. In the most
recent experiment \cite{Mundt2002}, the setup was similar but here the cavity
was resonant with an electric-quadrupole transition in Ca$^+$. This
transition was coherently driven by light coupled into the cavity, and the
excitation probability was found to be strongly modulated with the position
of the ion in the cavity mode.

All these studies are based on the interaction of a single atom with an
optical standing wave. While in the two Ca$^+$ experiments a standing wave of
resonant light forms inside the optical cavity, in the Ba$^+$ experiment part
of the electromagnetic mode structure (or vacuum field) around the ion is
transformed into a standing wave by the back-reflecting mirror.

Sub-wavelength resolution is achieved because the position of the ion
relative to the mirror(s) is well-controlled and the ion's spatial
wavefunction is confined to a region much smaller than the optical wavelength
$\lambda$ (400 to 800~nm). This strong confinement is due to the trapping
potential of a Paul-type ion trap, which to very good approximation can be
considered harmonic with typical oscillation frequencies $\Omega_t$ from 1 to
several MHz and a spatial extension of the lowest energy eigenstate around
10~nm. Laser cooling can prepare ions in thermal states with low average
quantum numbers. Preparation of the motional ground state with high purity
has also been achieved by means of sideband laser cooling
\cite{Diedrich1989,Monroe1995,Roos1999}. In the cases considered here, the
ions were Doppler-cooled, i.e.\ their motional energy $E_{th}$ was comparable
to the linewidth of the cooling transition which in all cases is about
20~MHz. Since the spatial extension scales with $\sqrt{\bar{n}}$, where
$\bar{n}=E_{th}/\hbar\Omega_t$ is the thermal energy in units of trap quanta,
a resolution between about 10 and 50~nm was achieved.

With this resolution, local variations of an optical field can be detected by
shifting the single ion through the field structure. Therefore this technique
received the name optical nanoscope \cite{Guthoehrlein2001}. Similar
sub-wavelength mapping techniques are used in microscopy \cite{Braun1998},
and they have also been demonstrated with single molecules instead of ions
\cite{Michaelis1999}.

While the typical length scale of an optical field is its wavelength, smaller
structures can emerge, e.g., in diffraction patterns, as high-order modes of
optical resonators, or in general when several partial waves are superimposed
and interfere. A standing wave is a comparatively simple structure, formed by
superposition of two counterpropagating travelling waves of equal amplitude
and polarization. It is nevertheless a highly instructive case because the
observation of a standing-wave structure is connected to the interference
between two processes pertaining to the two travelling waves. The resolution
with which the structure is detected determines the visibility of the
observed interference fringes. In turn, the observed visibility is a measure
for the resolution and hence for the spatial extension of the ion as well as
for its thermal energy.

On the other hand, a limited visibility in an interference measurement may be
a signature for the presence of which-way information which in principle can
be extracted from the system. In our cases, the which-way information must be
stored in the motional degrees of freedom of the ion, because it is the
motion in the trap which determines the spatial size and shape of the wave
packet of the ion.

The purpose of this paper is to present and compare these various physical
pictures and to show how thermal motion of the ion, resolution, visibility of
interference fringes, and which-way information are related. We will apply
these ideas to analyse the three mentioned experiments and compare the
experimental results. We will give classical and quantum descriptions of the
situations and show that, although their interpretations look rather
different, they are equally valid and lead to the same conclusions.

We focus here on the connection between ion motion, interference, and
resolution of the field structure. The question how the presence of a distant
mirror affects the \textit{internal} dynamics of the ion has been discussed
with a quantum model in Ref.~\cite{Dorner2002}.

\section{Interference and visibility}

\begin{figure*}[t]
\epsfig{file=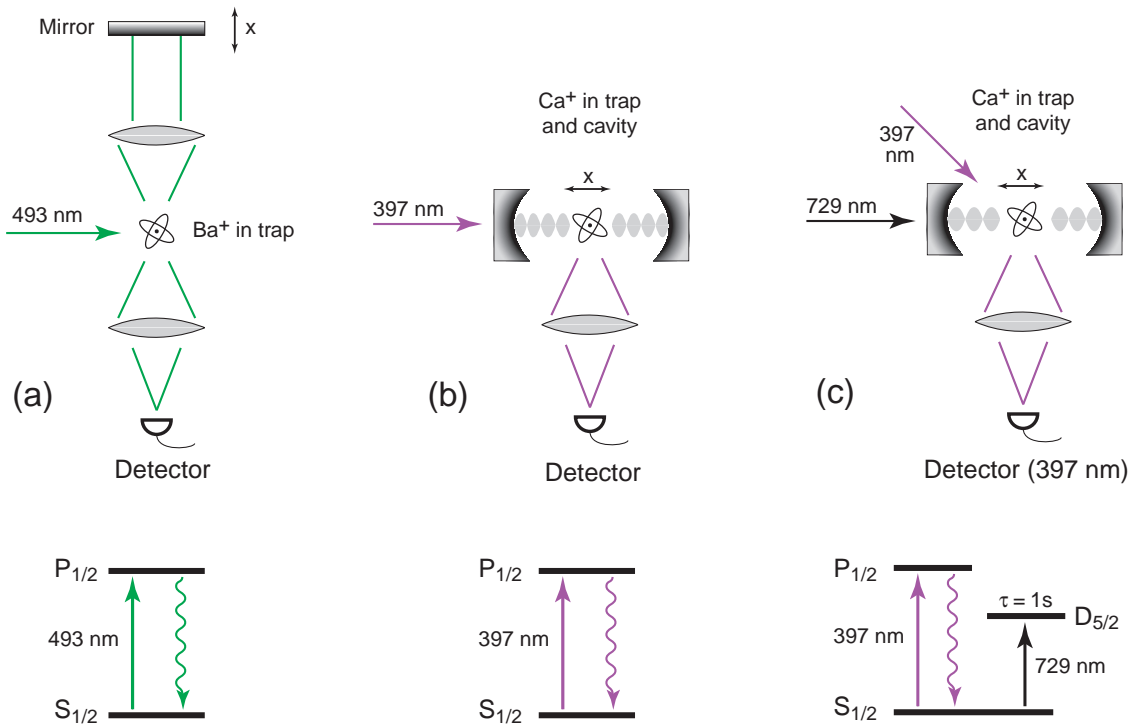, scale=0.8} 
\caption{\label{Sketch1} Schematics of the three experiments. (a)
Back-reflection experiment \cite{Eschner2001} with Ba$^+$, relevant levels
and laser wavelength. (b) Setup with ion scattering cavity light
\cite{Guthoehrlein2001}, relevant levels of Ca$^+$ and laser wavelength. (c)
Setup for quadrupole excitation by cavity light \cite{Mundt2002}, relevant
levels and lasers; after a pulse of cavity light (729~nm), the probability
for excitation into D$_{5/2}$ is measured through state-selective
fluorescence on the S$_{1/2}$ to P$_{1/2}$ transition (397~nm). In all
experiments the typical measurement time is much larger than the oscillation
period of the ion in the trap, $\Omega_t^{-1}$. }
\end{figure*}

First let us explain how in the three cases the observation of the standing
wave can be interpreted as an interference between two processes. In the
first experiment, a laser excites the ion from one side, and photons
scattered under 90$^{\circ}$ are detected, see Fig.~\ref{Sketch1}a. A mirror
is placed on the opposite side (at $-90^{\circ}$), such that photons
scattered into that direction are back-reflected and also sent into the
detector. Clearly the two pathways into the detector are indistinguishable
and interfere, which is observed as a modulation of the detector signal
(photon count rate) vs.\ the distance between mirror and ion
\cite{Eschner2001}. In the second experiment, where light from a cavity mode
is scattered by an ion \cite{Guthoehrlein2001}, two scattering amplitudes
corresponding to the two counterpropagating waves contribute to the detection
of a photon, see Fig.~\ref{Sketch1}b. Depending on the ion's position between
the cavity mirrors, these amplitudes are superimposed in or out of phase,
thus interfering constructively or destructively. The same explanation holds
for the cavity-induced quadrupole excitation \cite{Mundt2002}, only that here
the two excitation pathways into the long-lived upper state interfere, rather
than two scattering amplitudes.

A point-like probe would be able to map the interference fringes with perfect
visibility, but with a real atom the visibility will always be smaller. In
the qualitative discussion above, we have already used one of the possible
physical pictures for this visibility reduction: The spatial wave packet of
the ion probing the optical field structure acts as an apparatus function
with which the ideal signal (of a point-like probe) has to be convoluted to
find the experimental signal. This particular "spatial" picture applies most
intuitively to the cavity experiments. Taking a "temporal" view, the two
processes which interfere on the detector may indeed have happened (at the
ion) with a delay between them, during which the ion has moved. This is
certainly true for the mirror experiment where one partial wave is delayed by
the time it takes to the mirror and back, about 1.7~ns. The visibility will
be reduced because the two interfering pathways do not find the ion in
exactly the same state. There is also a "spectral" explanation for visibility
reduction: The ion oscillates in the trap so that, depending on the momentary
Doppler shift, it sees (or scatters) the two travelling waves with different
spectra. Only the overlapping spectral components can interfere.

Finally, there is a which-way interpretation. It takes into account that a
scattered or absorbed photon has a mechanical effect on the ion. On average,
every scattering event will leave one photon recoil in the atom, such that
its motional state may change in the course of the process. This recoil,
however, encodes with which of the two travelling waves the ion has
interacted, and inasmuch as this information is stored in the ion, the
visibility of the interference will be diminished.

We will now develop these different pictures in detail and compare the
conclusions to which they lead.

\section{Ion as apparatus function}

When we treat the wave packet of the ion as an effective apparatus function,
the observed signal is calculated as follows. Let $x$ denote the spatial
coordinate along the optical axis. The ideal signal from a point-like atom
probing the standing wave would be
\be \label{Ideal} S_{ideal}(x) = 2 \bar{S} \cos^2(kx) = \bar{S} \big(
1+\cos(2kx) \big)~, \ee where $\bar{S}$ is the average signal and
$k=2\pi/\lambda$ is the wave vector of the light. Now let $\rho(x-x_0)$ be
the probability to find the ion at position $x$ when the trap center is at
$x_0$. Then the observed signal as a function of the trap position $x_0$ is
\be \label{Observed} S(x_0) = \bar{S} \big (1+ V \cos(2kx_0) \big)~, \ee
where
\be \label{Visibility} V = \int_{-\infty}^{\infty}dx~ \rho(x) \cos(2kx) \ee
is the visibility of the interference fringes, in agreement with the standard
definition $V = (S_{max}-S_{min})/(S_{max}+S_{min})$. We have used that
$\rho$ is a symmetric function, which is certainly true for a harmonic trap.
In fact, as we will show later, in all cases which we treat here $\rho$ is a
Gaussian,
\be \label{Gauss} \rho(x) = \frac{1} {\sigma\sqrt{2\pi}}~
\exp(-\frac{x^2}{2\sigma^2}) \ee
with rms spatial extension $\sigma = \sqrt{\int_{-\infty}^{\infty} dx~x^2
\rho(x)}$. In this case the visibility according to Eq.~(\ref{Visibility}) is
\be \label{VisGauss} V = \exp(-2(k\sigma)^2)~. \ee

Relation (\ref{VisGauss}) is used to determine the resolution from a measured
visibility. The derived value of $\sigma$ will be an upper limit for the true
rms spatial extension of the ion, as other broadening effects may be present
in the experiment. The values for the three experiments, as calculated from
the measured visibility values $V=$~72\% \cite{Eschner2001}, 40\%
\cite{Guthoehrlein2001}, and 96.3\% \cite{Mundt2002}, are 32~nm, 43~nm, and
16~nm, respectively. It should be noted that in Ref.~\cite{Eschner2001} the
actual wave packet size is estimated to be 21~nm \cite{EschnerErr}, and the
reduced visibility is partly due to optical aberrations. It should also be
mentioned that the larger number given in Ref.~\cite{Guthoehrlein2001},
60~nm, is based on a different definition (by a factor $\sqrt{2}$) of the
spatial extension \cite{Lange2002} and is consistent with our result.

\section{Doppler effect}

In the case of the ion in front of a mirror, another description is
particularly intuitive which accounts for the time-dependent Doppler effect
of the ion oscillating in the trap. This view highlights the role of both the
delay between the two partial waves before they reach the detector, and of
the spectral effect of the ion's motion.

First we assume the ion to be oscillating classically with frequency
$\Omega_t$ and amplitude $x_c$, i.e.\ its momentary position is $x(t) = x_0+
x_c \sin(\Omega_t t)$. The oscillation modulates the phases of the two
partial waves $E_{\pm}$ emitted towards the detector (one directly, the other
via the mirror) according to
\be \label{Eplusminus} E_{\pm}(t) = E_0~ \textrm{e}^{i(kx_0 \pm kx_c
\sin(\Omega_t t)-\omega t)}~. \ee
These two fields reach the detector with a phase delay $e^{2ikL}$ between
them, where $L$ is the distance between trap center and mirror. The resulting
detector signal is
\be S_{c} = \label{TwoFields} |E_0|^2 \langle |\textrm{e}^{ikx_c
\sin(\Omega_t t)} + \textrm{e}^{2ikL} \textrm{e}^{-ikx_c \sin(\Omega_t t)}
|^2 \rangle~, \ee
where $\langle~\rangle$ denotes time averaging over many periods of the trap
oscillation. Such an integration time $T \gg \Omega_t^{-1}$ is used in all
the experiments discussed and will be assumed throughout our considerations.
From Eq.~(\ref{TwoFields}) we get the visibility reduction due to sinusoidal
oscillation
\be \label{ClassicalSignal} S_{c} = \bar{S}~ \big( 1+ J_0(kx_c) \cos(2kL)
\big)~, \ee
where $J_0$ is the zero-order Bessel function.

A laser-cooled ion is not oscillating classically but in a thermal state,
i.e. its oscillation amplitude $x_c$ follows a thermal probability
distribution. This distribution is derived from the Boltzmann distribution
for the ion's energy and is given by
\be \label{x_cDistribution} P(x_c)~dx_c = \frac{x_c}{\sigma^2}~
\exp(-\frac{x_c^2}{2\sigma^2})~ dx_c~, \ee
where $\sigma$ is the rms spatial extension as before, related to the thermal
energy by $E_{th} = M \Omega_t^2 \sigma^2$ with ion mass $M$. Combining
Eqs.~(\ref{ClassicalSignal}) and (\ref{x_cDistribution}) we get for the
detector signal
\be S(L) = \bar{S}~ \big (1+ \exp(-2(k\sigma)^2) \cos(2kL) \big)
\label{VisDoppler} \ee
in agreement with the previous result, Eq.~(\ref{VisGauss}). One finds the
same result when one evaluates first the spatial probability distribution for
the classical oscillator, $P_c(x) = (\pi \sqrt{x_c^2-x^2})^{-1}$ and
integrates it with distribution (\ref{x_cDistribution}), which yields the
Gaussian of Eq.~(\ref{Gauss}).

The picture of a Doppler effect is equally valid for the light scattering
from a cavity mode. In this case the phase delay $e^{2ikL}$ between the two
partial waves in Eq.~(\ref{TwoFields}) is replaced by the relative phase
between the two counterpropagating waves in the cavity, which varies as
$e^{2ikx_0}$ with the position $x_0$ of the trap center in the standing wave
($x_0=0$ is set to an antinode). In the same manner it applies to the
cavity-induced excitation. There Eq.~(\ref{TwoFields}) is interpreted as the
excitation probability when atomic saturation effects are neglected,
c.f.~Ref.~\cite{Mundt2002}.

Thus the phase modulation through the Doppler effect and the spatial
apparatus function are in fact only different pictures for the same
situation, yielding in all cases the same results for the visibility and the
resolution.

\section{Which-way information}

The relation between fringe visibility and which-way information in an
interference experiment is at the core of wave-particle duality. It has been
the subject of several general studies \cite{Englert1996}, and it was
recently studied in experiments with atom interferometers
\cite{Duerr1998,Bertet2001}.

With a trapped ion probing an optical field, the encoding of which-way
information happens through the recoil of an absorbed or an emitted photon.
To illustrate this, we will use the example of Fig.~\ref{Sketch1}b where an
ion scatters cavity light; later in this section we will show that the same
description applies to the other two experiments.

First assume that before scattering the ion is at rest. Depending on the
travelling wave from which a photon is absorbed, the photon recoil will leave
the ion oscillating with a certain initial momentum, i.e.\ a certain phase.
It is this phase which carries the which-way information. The second part of
the scattering process, the emission of the photon into the detector, will
always leave the same recoil kick and not introduce any further
distinguishability.

Now if every absorbed photon kicked the atom, the two final states pertaining
to the two travelling waves would always be different, and there would be no
interference. This is the extreme case of a very shallow trap ($\Omega_t
\rightarrow 0$), where the two possible recoil momenta accelerate the ion to
opposite sides, such that the two processes could be distinguished with
certainty. Because of the trapping potential, however, a certain fraction of
all absorption processes will leave the motional state unchanged. This is the
so-called Lamb-Dicke effect \cite{Dicke1953}, an important concept in laser
cooling of trapped atoms \cite{Stenholm1986}. It is this fraction of
scattering events which creates the interference.

In more detail, the recoil of the two travelling waves is transferred to the
ion's motional state by the spatial part of the respective electric field
operators, $\textrm{e}^{\pm ikx}$ \cite{RWA}. An initial energy eigenstate
$\ket{n}$ is thereby transformed into a superposition of states according to
\be \ket{n}~ \rightarrow~ \textrm{e}^{\pm ikx}\ket{n}~. \label{Kick} \ee
The overlap $\mean{n| \textrm{e}^{2ikx} |n}$ of these two possible final
states is expected to determine the visibility of the interference.

Using Eq.~(\ref{Kick}), the rate at which an ion would scatter photons from
one single travelling wave into the detector is given by \cite{Itano1998}
\be S_{RW} = S_{rest} \sum_n P(n) \sum_{k} |\mean{k|\textrm{e}^{\pm
ikx}|n}|^2~, \label{TravellingWave} \ee
where $S_{rest}$ is the scattering rate for an ion at rest, and $P(n) =
\bar{n}^n/ (\bar{n}+1)^{n+1}$ is a Boltzmann distribution over the harmonic
oscillator states \cite{Stationary}. For a standing wave, the matrix element
in Eq.~(\ref{TravellingWave}) is replaced by $\mean{k|2\cos(k(x-x_0))|n}$,
where $x_0$ is the position of the ion relative to an antinode as before.
This yields for the scattering rate from the standing wave, as a function of
the ion's position,
\be S(x_0) = 2 S_{rest} \Big( 1+ \sum_n P(n) \mean{n|\cos(2k(x-x_0))|n}
\Big)~. \ee
Since the spatial eigenfunctions $\ket{n}$ have definite parity, and using
$2S_{rest}=\bar{S}$, we get the interference signal
\be S(x_0) = \bar{S} \big( 1+ V\cos(2kx_0) \big)~, \ee where the visibility
is given by
\be V = \sum_n P(n) \mean{n|\cos(2kx)|n}~. \label{VisQuantum} \ee
Since $\mean{n|\cos(2kx)|n} = \mean{n|\textrm{e}^{2ikx}|n}$, we find that the
visibility of the interference fringes is indeed equal to the (thermally
averaged) overlap of the two possible final states $\textrm{e}^{\pm
ikx}\ket{n}$ of the individual processes, just as the which-way
interpretation suggests \cite{Classical}.

The same arguments are readily applied to the other two experiments: In the
case of Fig.~\ref{Sketch1}c, the photon recoil enters in exactly the same
way, only the signal $S(x_0)$ describes the position-dependent transition
probability into the upper state. In the case displayed in
Fig.~\ref{Sketch1}a, it is the recoil of the \textit{emitted} photons which
encodes the which-way information according to their direction of emission,
while absorption always happens from the same travelling-wave laser beam and
has no further effect.

Finally, we can evaluate Eq.~(\ref{VisQuantum}) using the properties of the
harmonic oscillator eigenfunctions \cite{Abramovitz}, and we find
\be V = \exp(-2(k\sigma)^2)~ \label{VisQuantumGauss} \ee where $\sigma$ is
again the rms spatial extension of the ion, now calculated from the thermal
distribution over the quantum states, $\sigma^2 = \sum_n P(n) \mean{n|x^2|n}
= (2\bar{n}+1)\mean{0|x^2|0}$.

Result (\ref{VisQuantumGauss}) is in perfect agreement with
Eq.~(\ref{VisGauss}) for the classical apparatus function and
Eq.~(\ref{VisDoppler}) for the time-dependent Doppler effect. This confirms
that the which-way interpretation which accounts for the photon recoil is
indeed an equally valid physical picture for the situation in the three
experiments and that it leads to the same conclusions.

Eq.~(\ref{VisQuantum}), in analogy with Eq.~(\ref{Visibility}), can also be
read as the Fourier transform of the thermal wave packet,
\be V = \int_{-\infty}^{\infty}dx~ \Big( \sum_n P(n) |\psi_n(x)|^2 \Big)
\cos(2kx)~, \ee where $\psi_n(x) = \mean{x|n}$ is the spatial representation
of the $n^{\textrm{th}}$ harmonic oscillator eigenfunction. Comparison with
Eq.~(\ref{VisQuantumGauss}) confirms that the thermal spatial probability
distribution is a Gaussian, as we assumed earlier.

We would like to note that the dependence of the visibility on the extension
of the motional wave function, Eq.~(\ref{VisQuantumGauss}), implies a limited
visibility also for an atom in the motional ground state. This should be
experimentally observable, e.g.\ in the case of Fig.~\ref{Sketch1}c, when the
probing of the cavity field is combined with ground state cooling techniques.
The dependence of the visibility on the ground state extension for different
trapping strength would illustrate nicely the quantum limit of confinement of
an atom, and it would be another fundamental demonstration of Bohr's moving
slit experiment, similar to the work of Ref.~\cite{Bertet2001}.

\section{Conclusions}

We have presented several physical pictures for the probing of an optical
field structure by a single trapped atom, and investigated the factors that
limit the spatial resolution. We used three recent experiments as examples,
where a standing wave structure was detected by a single trapped ion.

The detection of a standing wave involves the interference between the two
absorption or scattering processes pertaining to the two travelling waves,
therefore the spatial resolution is connected to the visibility of the
interference fringes. We have given several different explanations how a
limitation of the visibility arises: The spatial probability distribution of
the trapped atom can be regarded to act as an apparatus function with which
the ideal, full-contrast signal is convoluted. The ion's oscillation in the
trap can also be considered to create periodically phase-modulated light
fields, of which only the unshifted components interfere. Finally, the
possible modification of the ion's motional state by the photon recoil of the
two travelling waves can be considered to encode which-way information in the
ion.

These seemingly different pictures lead to identical conclusions regarding
their effect on the visibility, which shows that they are indeed only
different interpretations of the same physical situation.

Our study is not at all limited to a standing wave. This simple case only
helps to highlight the relations of more general validity, between phase
modulation, spatial probability distribution, and in particular the
interpretation of the photon recoil as which-way information.

\vspace{0.5cm}

\noindent \textbf{Acknowledgements.} The author wishes to thank Giovanna
Morigi and Rainer Blatt for clarifiying discussions and very helpful
comments. This work was partially supported by the Austrian Science Fund
(FWF, project SFB15).


\begin{thebibliography}{}
\bibitem{RWA}
The counterrotating part of the real field operator is off-resonant and
therefore neglected (rotating wave approximation).
\bibitem{Stationary}
While in an idividual scattering or absorption event the motional state may
change, we can still assume a stationary thermal distribution. In the first
and second experiment this is justified by the ongoing laser cooling, in the
third case by the average over many measurements after identical preparation.
\bibitem{Classical}
The analogous result is obtained for the visibility in a classical 2-path
interferometer.
\end{thebibliography}
\end{document}